\documentclass[screen,sigconf]{acmart}

\AtBeginDocument{%
  }

\copyrightyear{2025. This is the author's version of the work. It is posted here for your personal use. Not for redistribution. The definitive Version of Record was published in \textit{CHI Conference on Human Factors in Computing Systems (CHI '25), April 26-May 1, 2025, Yokohama, Japan}}
\acmDOI{10.1145/3706598.3713277}
\acmYear{2025}
\setcopyright{none}
\setcctype{}
\acmConference[CHI '25]{CHI Conference on Human Factors in Computing Systems}{April 26-May 1, 2025}{Yokohama, Japan}
\acmBooktitle{CHI Conference on Human Factors in Computing Systems (CHI '25), April 26-May 1, 2025, Yokohama, Japan} 
\acmISBN{979-8-4007-1394-1/25/04}

\usepackage{enumitem}   
\usepackage{lscape}
\usepackage{rotating}
\usepackage{stfloats}

\begin{document}

\title{Show Me the Work: Fact-Checkers' Requirements for Explainable Automated Fact-Checking}

\author{Greta Warren}
\email{grwa@di.ku.dk}
\orcid{0000-0002-3804-2287}
\affiliation{%
  \institution{Department of Computer Science, University of Copenhagen}
  \city{Copenhagen}
  \country{Denmark}
}
\author{Irina Shklovski}
\email{ias@di.ku.dk}
\orcid{0000-0003-1874-0958}
\affiliation{%
  \institution{University of Copenhagen}
  \city{Copenhagen}
  \country{Denmark}\\
    \institution{Linköping University}
  \city{Linköping}
  \country{Sweden}
}
\author{Isabelle Augenstein}
\email{augenstein@di.ku.dk}
\orcid{0000-0003-1562-7909}
\affiliation{%
  \institution{Department of Computer Science, University of Copenhagen}
  \city{Copenhagen}
  \country{Denmark}
}

\renewcommand{\shortauthors}{Warren et al.}

\begin{abstract}
  The pervasiveness of large language models and generative AI in online media has amplified the need for effective automated fact-checking to assist fact-checkers in tackling the increasing volume and sophistication of misinformation. The complex nature of fact-checking demands that automated fact-checking systems provide explanations that enable fact-checkers to scrutinise their outputs. However, it is unclear how these explanations should align with the decision-making and reasoning processes of fact-checkers to be effectively integrated into their workflows. Through semi-structured interviews with fact-checking professionals, we bridge this gap by: (i) providing an account of how fact-checkers assess evidence, make decisions, and explain their processes; (ii) examining how fact-checkers use automated tools in practice; and (iii) identifying fact-checker explanation requirements for automated fact-checking tools. The findings show unmet explanation needs and identify important criteria for replicable fact-checking explanations that trace the model's reasoning path, reference specific evidence, and highlight uncertainty and information gaps.
\end{abstract}

\ccsdesc[500]{Human-centered computing~Empirical studies in HCI}
\ccsdesc[300]{Human-centered computing~Empirical studies in collaborative and social computing}
\ccsdesc[300]{Computing methodologies~Natural language processing}

\keywords{Explainable AI, fact-checking, explanation, natural language processing, misinformation}


\maketitle

\section{Introduction} \label{intro}

The acceleration of misinformation and disinformation in recent years means that the role of fact-checkers in verifying public information continues to grow in scale and urgency \cite{adams2023misinformation}.
Fact-checking organisations have struggled to keep up with the increasing volume and sophistication of disinformation spread online \cite{arnold2020onlineFC,diakopoulos2020computational}. 
These concerns have been aggravated by the rise of generative AI.
For example, publicly available Large Language Models (LLMs) such as ChatGPT\footnote{\url{https://openai.com/chatgpt/}} enable the generation of text to facilitate the production of false news articles and information on social media at vast scale and speed \cite{augenstein2024factualityLLMs}, while recent developments in synthetic media generation enable production of convincing images, audio and video to mislead the public \cite{xu2023misinfogenai}.
The threat to societal stability posed by misinformation has bolstered calls for the development of AI-based tools to increase fact-checkers' capacity to debunk false claims, by fully or partially automating the fact-checking process \cite{Das2023state}.
This need becomes critical as a recent report indicates that the number of fact-checking projects worldwide has plateaued in recent years and decreased in 2024 \cite{Duke2024FactCheckingSputters}.
Despite the increasing amount of misinformation in the world requiring debunking, fact-checking projects are buckling under insufficient funding and political pressures \cite{IFCN2024report,MetaFactChecking2025}.
The development of Natural Language Processing (NLP) technologies for fact-checking is ramping up (see \cite{guo2022survey,Das2023state,kotonya2020explainablesurvey} for surveys), but their successful adoption relies on how effectively they can meet the needs of the fact-checkers they are intended for \cite{schlichtkrull2023usesfactchecking,langer2021we}.

Prior research has demonstrated that fact-checking is a complex endeavour requiring expert knowledge, research skills, and capacity to judge sources appropriately \cite{graves2018understanding,graves2017anatomy,micallef2022true}.
Fact-checkers themselves also report being broadly sceptical of AI tools and automation, believing that such technology is incapable of handling the intricacies and complexities of fact-checking (e.g., \cite{micallef2022true,juneja2022human,nakov2021automated}).
Previous work notes that fact-checker scepticism towards automated fact-checking tools could be alleviated if the tools could produce explanations and rationale for their outputs \cite{micallef2022true}.
Yet, it is unclear what specific information these tools must provide to be truly useful and how fact-checker demands for explainability and transparency might align with technical capabilities.
The fact-checking process tends to have discrete stages with different goals, comprising different tasks \cite{graves2017anatomy,juneja2022human,micallef2022true}. 
We speculate that at each of these stages, different types of automation and explanations may be required. 
However, little research has focused on how technical tools might fit into fact-checker processes and how explanations might assist and support fact-checkers in decision-making.
To address this gap we conducted an interview study with 10 fact-checking professionals from five continents, addressing the following research questions:

\begin{enumerate}
    \item \textbf{RQ1:} What factors do fact-checkers consider important in explaining their decisions and processes?
    \begin{itemize}
        \item What judgements do fact-checkers make when searching for claims, retrieving and assessing evidence, and making decisions based on evidence?
        \item What factors do fact-checkers consider when explaining their fact-checking decisions?
    \end{itemize}
        
\item \textbf{RQ2:} For which parts of the fact-checking processes are explanations of automated fact-checking systems useful?
\begin{itemize}
        \item For which parts of the fact-checking process are explanations necessary?
        \item What sort of information should explanations of automated fact-checking systems contain?
    \end{itemize}

\item \textbf{RQ3:} How can automated explanations of fact-checking decisions address the explanatory needs of fact-checkers?
\begin{itemize}
        \item What are the properties of a good explanation of an automated fact-checking system? 
        \item What contextual factors influence the utility of automated fact-checking explanations?
    \end{itemize}
\end{enumerate}
Our paper makes the following contributions:
(i) we provide an account of how fact-checkers select claims for verification, assess evidence, decide verdicts, and explain their processes;
(ii) we characterise how fact-checkers use automated tools in their workflow;
(iii) we identify the explanation needs of fact-checkers at each step of the fact-checking process; and
(iv) we highlight gaps between the current state of automated fact-checking and the practical needs of fact-checkers \textcolor{black}{and make recommendations for addressing these.}


\section{Related work} \label{relwork}

\subsection{The structure and practice of fact-checking} \label{relwork:factchecking}
Fact-checking has its origins in journalism and has emerged as a distinct practice in the last 20 years, spurred on by the proliferation of news through online and social media, and increased political polarisation \cite{smith2004factcheckersbible,graves2019fact,amazeen2020journalistic}.
Current literature distinguishes between
\textit{external} fact-checking, practised by independent fact-checking organisations 
(e.g., Snopes,\footnote{\url{https://www.snopes.com/}} Stowarzyszenie Demagog,\footnote{\url{https://demagog.org.pl/}} BOOM\footnote{\url{https://www.boomlive.in/}}), which involves analysing and verifying public claims such as those made in political statements, news reports, and on social media and
\textit{internal} fact-checking, practised by traditional news organisations, cross-checking and correcting other journalists' work before publication to filter inaccuracies and to protect the publisher from potential liabilities
\cite{graves2019fact,juneja2022human}.
In this work, we focus on the needs of independent (external) fact-checkers, although we anticipate that our findings will be applicable to internal fact-checkers, whose work tasks typically comprise a subset of the fact-checking tasks.
The practice of fact-checking encompasses close collaboration between fact-checkers and news editors who supervise the work, copy editors responsible for the quality of fact-checks, investigators and researchers with expertise in data analysis and visualisation tools, and social media managers who disseminate and maximise impact and engagement of fact-checks \cite{juneja2022human}. 
Fact-checkers are typically tasked with a wide remit of duties, such as monitoring social media, fielding reader requests, extracting and prioritising claims, researching, consulting data and domain experts, assigning veracity labels, and writing up fact-checks \cite{juneja2022human, micallef2022true, graves2017anatomy}.
Ethnographic and interview studies have documented how fact-checkers worldwide follow broadly similar processes \cite{graves2017anatomy,micallef2022true,juneja2022human,koliska2024epistemology}.
Previous work has grouped these processes into four steps that comprise an archetypal fact-checking pipeline: (i) choosing claims to check; (ii) searching for evidence; (iii) assigning a verdict; and (iv) writing and communicating the fact-check \cite{Das2023state,guo2022survey,micallef2022true,juneja2022human}.\footnote{Some studies (e.g., \cite{micallef2022true,juneja2022human}) also include the additional step of publishing and disseminating the fact-check in this pipeline, however, as this task is usually performed by social media managers rather than fact-checkers, we focus on the four listed here.}

Previous qualitative studies have provided rich accounts of fact-checkers' workflows and the challenges they face \cite{micallef2022true}, detailing the human and organisational infrastructures and stakeholder groups that underlie them \cite{juneja2022human}, their core values \cite{dierickx2023automated} and tensions between their epistemological ideals and practices \cite{bengtsson2024rhetoricalfactcheckers}. However, the precise decision-making and reasoning processes employed by fact-checkers in their work remain unclear.
Studies frequently describe the antipathy of fact-checkers towards integrating automated fact-checking techniques in their work, referring to their lack of utility in practice and the absence of explanations provided by opaque systems \cite{micallef2022true,liu2023humancenteredNLP}.
\textcolor{black}{Yet existing literature does not present tangible solutions for model developers regarding how such flaws can be addressed to support how fact-checkers reason and make decisions about the claims they check.}
We aim to develop a more fine-grained understanding of how fact-checkers evaluate information, assign verdicts, and explain these processes \textcolor{black}{(\textbf{RQ1})}, as a prerequisite to designing computational systems and tools that support and enhance decision-making.

\subsection{The state of automated fact-checking} \label{relwork:automated_FC}
Automated fact-checking systems tend to follow a fact-checking pipeline similar to the one described in the section above \cite{Das2023state,guo2022survey,dmonte2024claimverificationagelarge}. The primary tasks are: 
(i) claim detection and claim filtering \cite{hassan2017ClaimBuster,hassan2017TowardsClaimBuster}; 
(ii) evidence retrieval \cite{clarke2020overview,li2021anasearch};
(iii) veracity prediction \cite{alhindi2018evidence,thorne2018fever,augenstein2019multifc}; and (iv) explanation generation \cite{atanasova2020generating,kotonya2020fcpublichealth}.
Claim detection involves identifying checkable claims 
from sources such as social media, news articles or live political debates \cite{hassan2017ClaimBuster}. 
Successful techniques involve human-in-the-loop and semi-supervised active learning approaches where the user of a system can provide feedback or labels for selected instances \cite{farinneya2021activelearningrumor, tschiatschek2018fakenewsdetection}. Additional methods for claim filtering and prioritisation identify urgent claims (based on virality or harmfulness) \cite{nakov2022clef,adair2020automatedjournalism,wright-augenstein-2020-claim}.
Stance detection techniques \cite{augenstein-etal-2016-stance,ferreira2016emergent,popat2018declare} are used to classify whether a given piece of evidence supports or refutes a claim. 
Although veracity prediction is performed based on the retrieved evidence, the evidence retrieval step is typically conducted in a coarse-grained way using standard search engines, which are optimised for relevance rather than for veracity prediction, leading to subpar retrieval performance \cite{Das2023state,hardalov-etal-2022-survey,hagstrom2024realitycheckcontextutilisation}. Previous work in information retrieval (IR) has examined the challenge of retrieving credible and relevant information from online sources \cite{clarke2020overview}.
Automated veracity prediction techniques attempt to determine the veracity of a claim given provided evidence,
tending to rely on secondary evidence documents such as news articles \cite{pomerleau2017fakenews,ferreira2016emergent}, Wikipedia \cite{thorne2018fever} or retrieved online sources \cite{augenstein2019multifc}.
These models vary in the number of veracity labels they consider. Where some retain the original labels assigned by fact-checkers \cite{augenstein2019multifc}, others collapse categories such as ``true'' and ``mostly true'', or ``mixed evidence'' and ``unproven'' \cite{hanselowski2019corpus,kotonya2020fcpublichealth,gupta2021xfactmultilingual}.

A growing body of work has examined methods of generating explanations for the decisions of automated fact-checking systems.
Explainable fact-checking methods can be grouped into five categories.
Attribution-based explanations highlight aspects of the evidence (e.g., individual tokens or words) that contributed to a predicted verdict \cite{popat2018declare}.
Rule-based explanations provide a set of rules that describe parts of the decision-making process \cite{gad2019exfakt}.
Counterfactual or adversarial examples that identify minimal changes in an input that can alter a model's prediction provide insight to models' weaknesses or biases \cite{atanasova2020adversarial}.
Case-based explanations provide a rationale for a model's decision by showing how similar instances were assigned the same label by a human \cite{das2022prototex}.
Summarisation explanation methods provide natural language summaries of the evidence to demonstrate that it justifies the verdict \cite{atanasova2020generating,kotonya2020fcpublichealth}.
All these explanation methods have practical limitations.
For example, the validity and utility of feature attributions as explanations is the subject of ongoing debate (e.g., \cite{jain2019attention,bibal2022attentionexpl}), while rule, summarisation, and case-based explanations assume the existence of an existing knowledge base, a fact-checking article already written by a fact-checker, and human-annotated similar examples, respectively.
Hence, \textcolor{black}{current AI tools for} veracity prediction and explanation have limited utility \textcolor{black}{in real-world applications, such as}  when applied to unseen claims, for which there are no curated evidence documents or fact-checking articles.

\textcolor{black}{Not only are comprehensive empirical evaluations of automated fact-checking tools sparse, automated fact-checking research has been criticised for its disconnect from the practical realities of fact-checking. Existing evidence demonstrates significant gaps between the capabilities of available automated fact-checking tools and fact-checker needs.
For example, professional journalists, evaluating one proposed automated tool for veracity prediction, reported that only 59\% of the claims were accurately classified, and just 58\% of the evidence sentences retrieved were relevant to the claim \cite{miranda2019automated}.
A content analysis of automated fact-checking research articles found that they were often vague and inconsistent in linking proposed methods to their claimed purposes: only 31\% of papers with the stated aim of automating the fact-checking process included evidence retrieval methods, while 81\% relied solely on classification (or veracity prediction)  \cite{schlichtkrull2023usesfactchecking}.
}
Our work seeks to document the specific tasks where fact-checkers use automated fact-checking tools and where explanations of automated fact-checking are needed (\textbf{RQ2}) and investigate how automated fact-checking tools can better align with fact-checker requirements.


\subsection{Explainable AI for fact-checking} \label{relwork:explainableAI}

Various desiderata for explanations of AI systems have been proposed in recent years \cite{sokol2020explassessment,langer2021we,nauta2023xaireview}. Of these, researchers have stipulated eight criteria specific to fact-checking explanations \cite{kotonya2020explainablesurvey}. 
These criteria propose that fact-checking explanations should be actionable (i.e., provide steps towards desirable outcome), causal (i.e., use a causal model), coherent (i.e., follow natural laws, be rule-based or otherwise deterministic), context-full (i.e., presented in the context of the claim), interactive (i.e., allow users to provide feedback to the system), parsimonious (i.e., communicate necessary information with minimal redundancy), chronological (i.e., reflect when a statement was made and the information available at that time), and impartial (i.e., avoid partisan language or opinions).
While these criteria are useful for NLP researchers developing basic explanation techniques, an important caveat is that these criteria are what model developers intuit as satisfactory fact-checking explanation, rather than being grounded in the needs of the people that might use these explanations, such as fact-checkers using automated tools.
\textcolor{black}{This limitation is a further example of the disconnect between automated fact-checking research and real-world applications and illustrates a neglect of} stakeholder needs \cite{schlichtkrull2023usesfactchecking}, crucial for explanations to be truly useful \cite{liao2022humancenteredexplainableaixai,langer2021we}.
Automated fact-checking encompasses different groups of stakeholders (e.g., fact-checkers, content-moderators, model developers, and laypeople), each with distinct explanation needs, tailoring to \textcolor{black}{contextual factors}, goals, and levels of expertise \cite{langer2021we,juneja2022human,liao2022humancenteredexplainableaixai,ehsan2024thewho}.
However, existing empirical evaluations of explainable fact-checking are almost exclusively directed at and executed with laypeople \textcolor{black}{from a limited selection of Western countries}, rather than expert fact-checkers \textcolor{black}{with diverse and varied contexts and perspectives}.
\textcolor{black}{One such study indicated that neither feature-attribution nor example-based explanations of automated veracity prediction had an effect on laypeople's perceptions of the veracity of a news story or their intent to share it, but increased their tendency to over-rely on the AI system when it provided incorrect predictions \cite{lim2023xai}. A separate study also found no effect of example-based explanations on people's accuracy in predicting the veracity of a claim \cite{linder2021level}.}
Illustrating that current explanation methods have little utility for fact-checkers, a recent study found that free-text explanations for an automated disinformation detection system improved the performance of laypeople in identifying false information, but not those of journalists \cite{schmitt2024explhuman}.
\textcolor{black}{Together, these studies suggest that existing explainability techniques are, at best, ineffective and, at worst, misleading and vulnerable to misuse, potentially bolstering misinformation instead of dispelling it.
Both empirical \cite{schmitt2024explhuman} and literature-based \cite{schlichtkrull2023usesfactchecking} analyses indicate that automated explainable fact-checking research} is shaped by researcher assumptions rather than by direct engagement with fact-checkers, which hinders their utility \cite{kotonya2020explainablesurvey,Das2023state}.
\textcolor{black}{We seek to address these shortcomings by identifying the explanation needs of fact-checkers and how they can be addressed by explainable fact-checking systems (\textbf{RQ3}}).

\begin{table*}[h]
    \centering
    \begin{tabular}{llllll}
\hline
\textbf{Participant ID} & \textbf{Country} & \textbf{Occupation}                  & \textbf{\begin{tabular}[l]{@{}l@{}}Organisational \\ Context\end{tabular}} & \textbf{\begin{tabular}[l]{@{}l@{}}Fact-Checking \\ Experience\end{tabular}} & \textbf{Gender} \\ \hline
P1                      & Ukraine          & Investigative Journalist             & Freelance                                                                  & 8 years                                                                      & Female          \\
P2                      & Argentina        & Fact-Checker                         & Independent                                                                & 4 years                                                                      & Female          \\
P3                      & Poland           & Fact-Checker                         & Independent                                                                & 4 years                                                                      & Male            \\
P4                      & USA              & Investigative Journalist  \& Trainer & Freelance                                                                  & 12 years                                                                     & Female          \\
P5                      & Poland           & Fact-Checker                         & Independent                                                                & 5 years                                                                      & Female          \\
P6                      & Ireland \& USA   & Fact-Checker \& Project Manager      & Independent                                                                & 4 years                                                                      & Male            \\
P7                      & Poland           & Director \& Journalist               & Independent                                                                & 4 years                                                                      & Male            \\
P8                      & Zimbabwe         & Fact-Checker                         & Independent                                                                & 2 years                                                                      & Female          \\
P9                      & Nigeria          & Investigative Journalist             & Independent                                                                & 4 years                                                                      & Male            \\
P10                     & India            & Fact-Checker                         & Independent                                                                & 6 years                                                                      & Male            \\ \hline
\end{tabular}
    \caption{Demographics of the interview participants}
    \label{tab:participants}
    \Description["Demographics of the interview participants" presents information about ten participants, focusing on their Participant ID, Country, Occupation, Organisational Context, Fact-Checking Experience, and Gender. The data for each participant is as follows:
P1 (Ukraine) - Investigative Journalist, Freelance, 8 years, Female
P2 (Argentina) - Fact-Checker, Independent, 4 years, Female
P3 (Poland) - Fact-Checker, Independent, 4 years, Male
P4 (USA) - Investigative Journalist \& Trainer, Freelance, 12 years, Female
P5 (Poland) - Fact-Checker, Independent, 5 years, Female
P6 (Ireland \& USA) - Fact-Checker \& Project Manager, Independent, 4 years, Male
P7 (Poland) - Director \& Journalist, Independent, 4 years, Male
P8 (Zimbabwe) - Fact-Checker, Independent, 2 years, Female
P9 (Nigeria) - Investigative Journalist, Independent, 4 years, Male
P10 (India) - Fact-Checker, Independent, 6 years, Male]
    
\end{table*}

\section{Method} \label{method}
The central goal of this research is to explore fact-checking as a primary use case for explainable NLP systems, through a collaboration between NLP and Human-Computer Interaction (HCI) researchers.
We jointly developed new questions and adapted those used in prior work on fact-checkers \cite{micallef2022true} for our interview guide for semi-structured interviews with professional fact-checkers and journalists (see Appendix \ref{app:int_questions}).
To elicit discussion of fine-grained decision-making, we asked fact-checkers to describe their processes and reasoning by referencing a claim that they had worked on recently \textbf{(RQ1)}.
For each step in the process, we asked participants about their use of automated fact-checking tools and their effectiveness \textbf{(RQ2)}.
To uncover specific explanation needs for automated fact-checking tools, we asked how fact-checkers explained their own decisions and processes and about their understanding of how their automated tools worked \textbf{(RQ3)}.
Follow-up questions were guided by the interviewees' expertise; for instance, participants with more experience using AI-based tools were asked in more detail about how these worked and how they selected certain tools over others.
We also adapted the interviews to pursue developing themes. For example, we noticed that fact-checkers were sensitive to confidence scores output by AI tools, so we asked them how they understood such expressions of uncertainty and the strategies they used to resolve them.
The study was approved by our institution's Research Ethics Committee (reference number 504-0516/24-5000).

\subsection{Participant recruitment} \label{method:participants}
We recruited 10 fact-checking professionals (see Table \ref{tab:participants} for an overview of demographics)
by advertising the study on mailing lists for fact-checkers and investigative journalists (the International Fact-Checking Network (IFCN)\footnote{\url{https://www.poynter.org/ifcn/}} and Hacks/Hackers\footnote{\url{https://www.hackshackers.com/}}), social media websites (Twitter/X and LinkedIn), and at an international conference for professional fact-checkers (\hyperlink{https://www.globalfact11.com/event/0116938c-093b-4f34-afb7-09c3312aaa98/summary}{GlobalFact11}) in June 2024. 
Our recruitment material consisted of a brief description of the study, the research team, a link to a webpage containing more detailed information and an online registration form.

In line with prior studies \cite{micallef2022true}, recruiting fact-checkers proved challenging.
The interviews took place in June and July 2024, during which several major election campaigns (e.g., the European Parliament elections, Indian general election, United States presidential nominations) took place, which may have exacerbated recruitment challenges.
We received a large volume of spurious responses (>200), which were identified by rapid form-completion times, invalid links to professional portfolios, and incoherent or inappropriate responses to open-ended questions. 
After filtering spam responses, we received 31 genuine expressions of interest.
We emailed these respondents with information about the study, the information sheet, and an informed consent form. Participants returned the informed consent form, scheduled an interview slot with the first author and filled out a short pre-interview questionnaire (Appendix \ref{app:pre-int}).

Our sample comprised 5 female and 5 male fact-checkers from Europe, Africa, Asia, North America and South America.
Eight participants worked for fact-checking organisations, while two (P1 and P4) were freelance investigative journalists who had previously worked for independent fact-checking organisations for at least 8 years.
Three participants (P6, P7, and P9) worked for fact-checking organisations where they held managerial and editorial roles in addition to fact-checking work.
\textcolor{black}{While our participant sample was geographically diverse, previous work has documented that fact-checkers worldwide share common objectives and practices \cite{micallef2022true,juneja2022human}.
In keeping with prior studies, our interviewees followed well-defined and consistent work processes, demonstrating many commonalities in the use of automated tools.
The diversity of our sample enriched our findings by broadening the range of perspectives and contexts foregrounded in the current work, by providing insight into constraints that fact-checkers adapted to when using automated fact-checking tools, as a result of regional (e.g., poor AI tool performance on non-Western languages and accents) and resource-based (e.g., reliance on free or open-source tools) inequities.
Our research questions were focused and relatively narrow, and we reached theoretical saturation after eight interviews, at which point no new themes developed from our analyses. We conducted two additional scheduled interviews, which confirmed this.}

\subsection{Data collection} \label{method:data_collection}
All interviews were conducted remotely via Zoom and lasted approximately 1 hour. 
Conversations focused on (i) how fact-checkers describe and explain their work processes, (ii) their use of AI and technological tools in their work, and (iii) their information and explanation needs in automated fact-checking (Appendix \ref{app:int_questions}).
All interviews were recorded and transcribed with participant permission.
Participants were compensated with a USD \$50 online gift voucher.

\subsection{Data analysis} \label{method:data_analysis}

We analysed the interview transcripts using an iterative bottom-up, inductive approach inspired by grounded theory \cite{corbin2015basics}, \textcolor{black}{using NVivo 14 software}.
The first author coded the transcripts using line-by-line open coding, \textcolor{black}{(e.g., "triangulate verdicts", "using multiple tools for diverse results").} 
Axial coding (e.g., "multi-tool usage") was used to further develop the themes from the data. 
\textcolor{black}{These codes were reviewed after analysing the first two transcripts and refined accordingly. For example, some participants reported using multiple AI tools to resolve conflicting model uncertainty scores, which was initially assigned to the "multi-tool usage" theme. However, after pursuing this theme in later interviews, it became clear that fact-checkers' demand for explanations of AI uncertainty was sufficiently distinctive and developed to merit a stand-alone theme.}
The resulting codes
were discussed by all authors and categorised into higher-level concepts \textcolor{black}{using selective coding, such as "AI tools for veracity prediction".}
We collaboratively iterated over these concepts to identify relevant themes in the data \textcolor{black}{and map them to our research questions.}
\textcolor{black}{While there were few outright disagreements during this process, the distinct perspectives of the authors (a team of NLP and HCI researchers) engendered debate and discussion about the technical challenges of addressing the requirements of fact-checkers, as well as fundamental differences in how automated fact-checking is approached in NLP research (e.g., often in artificial set-ups with the aim of improving model performance on benchmark datasets \cite{schlichtkrull2023usesfactchecking}) compared to fact-checking `in the wild'.
These discussions shaped our aim to identify lacunas in NLP and HCI research with potential for concrete improvements in automated tools for fact-checkers. 
}
Table \ref{tab:RQ_themes} displays the main themes related to our research questions developed through our analysis.
\textcolor{black}{More detailed examples of the subthemes and codes are available in Appendix \ref{app:themes_codes}.}

\begin{table}[h]
    \centering
    \begin{tabular}{l l}
    \hline
        \textbf{Research Questions} & \textbf{Themes} \\
        \hline
        \textbf{RQ1:} What factors do & Evidence quality (\textbf{10}) \\ fact-checkers consider & Deciding verdicts (\textbf{10})\\ important in explaining & Verdict-dependent (\textbf{6}) \\ their decisions and & Communicating complexity (\textbf{9}) \\ processes?
        \\
        \\
         \textbf{RQ2:} For which parts of & Claim detection (\textbf{6}) \\ the fact-checking process & Evidence retrieval (\textbf{8}) \\ are explanations of & Veracity prediction (\textbf{8}) \\ automated fact-checking & Communicating fact-checks (\textbf{4})
         \\
         systems useful? \\
         \\
         \textbf{RQ3:} How can explanations & Explain processes (\textbf{9}) \\
         of fact-checking decisions & Replicability (\textbf{9}) \\ address the explanatory & Explain uncertainty (\textbf{8}) \\
         needs of fact-checkers? & Verifiability (\textbf{9}) \\
         & Faithfulness (\textbf{3}) \\
    \hline    
    \end{tabular}
    \caption{Key themes relevant to each research question, with the number of participants who mentioned each theme in parentheses.}
    \label{tab:RQ_themes}
    \Description[Key themes relevant to each research question" presents three research questions (RQ) and the corresponding themes, along with the number of participants who mentioned each theme]{Key themes relevant to each research question" presents three research questions (RQ) and the corresponding themes, along with the number of participants who mentioned each theme (shown in parentheses).
    RQ1: What factors do fact-checkers consider important in explaining their decisions and processes?
    Themes:
    Evidence quality (10 participants)
    Deciding verdicts (10 participants)
    Verdict-dependent (6 participants)
    Communicating complexity (9 participants)
    RQ2: For which parts of the fact-checking processes are explanations of automated fact-checking systems useful?
    Themes:
    Claim detection (6 participants)
    Evidence retrieval (8 participants)
    Veracity prediction (8 participants)
    Communicating fact-checks (4 participants)
    RQ3: How can explanations of fact-checking decisions address the explanatory needs of fact-checkers?
    Themes:
    Explain processes (9 participants)
    Explain uncertainty (8 participants)
    Replicability (9 participants)
    Verifiability (9 participants)
    Faithfulness (3 participants)}
\end{table}

\subsection{Methodological limitations} \label{method:limitations}

Similar to prior research \cite{micallef2022true} our sample size is small, given the relatively small global population of fact-checkers.\footnote{As of May 2024, there were 439 independent professional fact-checking projects in 111 countries \cite{Duke2024FactCheckingSputters}}
Furthermore, we only conducted interviews in English, and though our sample was geographically diverse and contained many non-native English speakers, this constraint may have limited participation.
Our findings are also subject to the inherent subjectivity associated with self-reported data \cite{donaldson2002understanding}.

\section{Findings} \label{findings}
\begin{figure*}[!t]
    \centering
    \includegraphics[width=\textwidth]{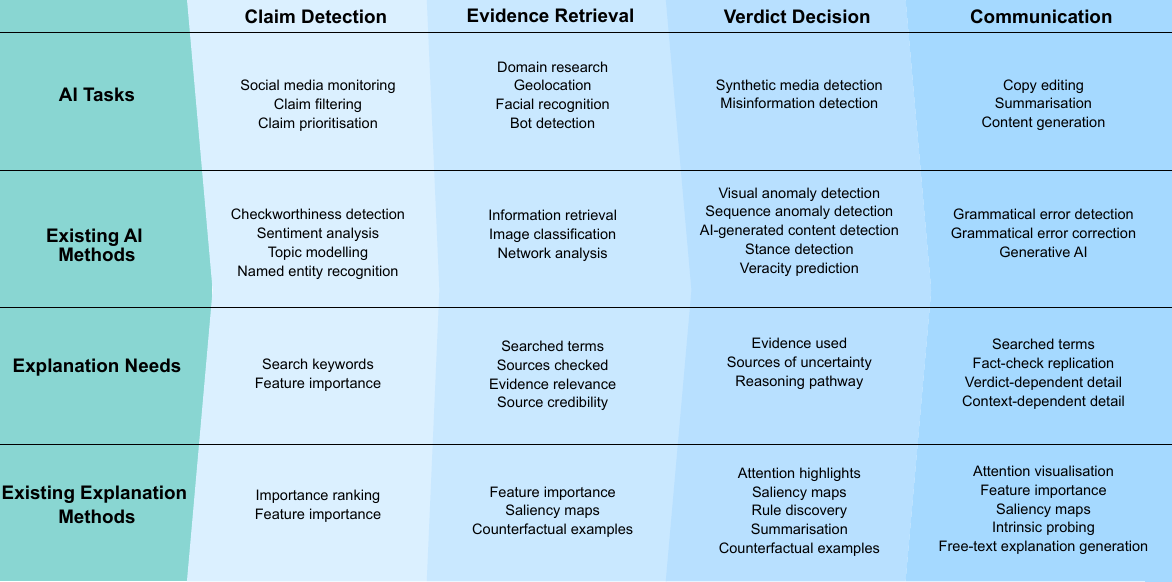}
    \caption{Description of fact-checkers' AI tool use and explanation needs, contrasted with what AI methods and explanation methods have been researched}
    \Description[A flowchart depicting four stages in the automated fact-checking process: Claim detection, Evidence retrieval, Verdict decision, and Communication. Each stage lists AI tasks, existing AI methods, explanation needs, and existing explanation methods.]{A flowchart depicting four stages in the automated fact-checking process: Claim detection, Evidence retrieval, Verdict decision, and Communication. Each stage lists AI tasks, existing AI methods, explanation needs, and existing explanation methods.
    1. Claim detection:
    * AI tasks: Social media monitoring, claim filtering, claim prioritisation.
    * Existing AI methods: Checkworthiness detection, sentiment analysis, topic modeling, named entity recognition.
    * Explanation needs: Search keywords, feature importance.
    * Existing explanation methods: Importance ranking, feature importance.
    2. Evidence retrieval:
    * AI tasks: Domain research, geolocation, facial recognition, bot detection.
    * Existing AI methods: Information retrieval, image classification, network analysis.
    * Explanation needs: Evidence relevance, source credibility, sources checked, search terms.
    * Existing explanation methods: Feature importance, saliency maps, counterfactual/adversarial examples.
    3. Verdict decision:
    * AI tasks: Synthetic media detection, misinformation detection.
    * Existing AI methods: Visual anomaly detection, sequence anomaly detection, detection of AI-generated content, stance detection, veracity prediction.
    * Explanation needs: Evidence used, sources of uncertainty, line of reasoning.
    * Existing explanation methods: Attention highlights, rule discovery, summarization, counterfactual/adversarial examples, saliency maps.
    4. Communication:
    * AI tasks: Copy editing, summarization, content generation.
    * Existing AI methods: Grammatical error detection and correction, generative AI.
    * Explanation needs: Fact-check replication, verdict and context-dependent detail.
    * Existing explanation methods: Attention visualization, feature importance, saliency maps, intrinsic probing, free-text explanation generation.
    Green arrows show the flow from one stage to the next, indicating how the output of one stage feeds into the next in the fact-checking process. Each column is uniformly structured, with explanations of tasks, AI methods, and the types of explanations needed at each step.}
    \label{fig:expl_needs}
\end{figure*}

We set out to document specific work practices of fact-checkers related to evaluating claims and explaining their decisions in greater detail than has been previously documented \textcolor{black}{(\textbf{RQ1})}.
We also aimed to investigate fact-checker use of automated fact-checking tools \textcolor{black}{(\textbf{RQ2})} to identify specific information and explainability criteria tailored to automated fact-checking \textcolor{black}{(\textbf{RQ3})}.
For each step of the fact-checking process \cite{Das2023state,guo2022survey}, Figure \ref{fig:expl_needs} depicts the type of AI tools used by fact-checkers, the AI methods underlying them, the explanation needs expressed by fact-checkers, and the existing methods that have been proposed in the explainability literature for the specific sub-task. 
From participant descriptions of their decision-making processes and tool use, we found that AI tool usage was present in each part of the fact-checking pipeline. In line with previous findings  
\cite{micallef2022true,Das2023state}, fact-checker tool use was fragmented; available tools had limited scope and performed one specific function (e.g., bot-detection for Twitter/X accounts), rather than automating multiple tasks.
We also found that the explanations fact-checkers required from AI tools differed depending on the fact-checking task, and that these explanation needs only partially aligned with explanation methods proposed by the literature.

\subsection{Claim detection} \label{findings:claim_detection}
\subsubsection{\textcolor{black}{Decision-making in claim detection}}
Participants described selecting claims from trending social media posts, such as TikTok, Facebook and Twitter/X, as well as receiving tips from the public via tiplines, which could be particularly useful for claims circulating on private messaging platforms such as WhatsApp.

P9, who had an editorial role, described the process of his organisation's weekly editorial meeting: "Journalists submit claims for likely verification, which I review before they jump on it to do their research [...] we discuss what are the trending claims, the ones that are fact-checkable. [Then] we assign them to our fact-checkers."

\subsubsection{Automated tools for claim detection}
Six participants (P2, P2, P6, P7, P9, P10) reported using AI tools for monitoring social media for detecting check-worthy and potentially harmful claims, such as Logically Accelerate\footnote{\url{https://www.logically.ai/accelerate}}, Rolli\footnote{\url{https://rolliapp.com/}}, 
CrowdTangle\footnote{CrowdTangle was discontinued by Meta in August 2024, and replaced by Meta Content Library (\url{https://transparency.meta.com/researchtools/other-datasets/crowdtangle})}, as well as Full Fact AI's tool\footnote{\url{https://fullfact.org/ai/about/}} for live debate claim-monitoring. Despite the proliferation of such tools, cost played a key role for fact-checkers coming from lower-resourced locations (P5, P7, P9, P10):
\begin{quote}
    "Unfortunately, we can't afford to pay for the premium features to use some of those [live debate claim detection] tools again. That's why we restrict ourselves to the open source that are freely available online" 
    --- P9, Investigative Journalist, Nigeria
\end{quote}
Yet paid tools, even where fact-checkers had resources to use them, created their own challenges. P5, P9, and P10 explained that using open-source tools allowed readers to replicate their fact-checks once the verdict and its explanation are released:
\begin{quote}
    "We are not using any AI recognition tools that are paid [...] because we want to give a chance to our readers to check our facts [...] we use those tools that they can use as well." 
    --- P5, Fact-Checker, Poland
\end{quote}

\subsubsection{Explanation needs for automated claim detection}
Participants were generally positive about tools used for claim detection, noting that they helped to expedite identification of potential checkable claims, reduce work required to monitor multiple media channels, and surface claims they may have otherwise missed.
Most reported not giving much thought to how these tools work, but those that did, tended to develop their own folk theories \cite{de2022invisible} about them. For example, P9 described his theories about how the Full Fact tool for live debate claim detection worked from observing patterns in the claims it detected:

\begin{quote}
    "Any claim that has the `most popular', `most especially', that has this kind of a clause, `highest', `lowest', `the best', the most this, the least this. I've seen it concurrently picking out words or phrases that has those adjectives. So that's how I [...] determine, `Okay, this is how this algorithm is working' [...] especially any sentences that has figures involved in them, it also flags a potential claim for us to verify." --- P9, Investigative Journalist, Nigeria
\end{quote}
Most participants were not interested in understanding how these tools worked, but appreciated what they perceived as support for a time-consuming task. Claim detection is a relatively low-stakes task where there are many tools, and thus, opportunities for triangulation.  Inaccuracies in a particular tool were not deemed a significant drawback \cite{Das2023state}. Where participants developed their own folk theories, such theories helped them decide when and how to use the tools and how to evaluate the delivered suggestions.
Of course, risks associated with automated tools for claim detection remain. For example, existing biases in claim selection (e.g., \cite{singh2024factcheckingpoliticalneutrality}) may be magnified, or new biases may be introduced in the claim selection process \cite{Dodge2019ExplainingJudgment}. 
Here, explanations for how and why particular claims are selected may be helpful, but capacity to coordinate the use of multiple tools and to address language deficiencies in tool performance may be more important.

\subsection{Evidence retrieval} \label{findings:evidence_retrieval}
\subsubsection{Decision-making in evidence retrieval}
For our participants, evidence retrieval was the core of the fact-checking task. 
The crucial aspect of this stage was retrieval of \textit{primary sources} for evidence, such as court documents, 
official statements, a person who appears to be the source of the claim, or photos and videos of an alleged event:
\begin{quote}
    "We don't cite, for example, media citing source, we try to find always the most primary, most original source of data [... If] we're talking about migration statistics, we try to look at the migration officials, maybe ask them for an official statement." --- P3, Fact-Checker, Poland
\end{quote}
P1 explained that she avoided citing secondary sources such as related news articles, because "media will provide interpretation".
Fact-checkers also emphasised the importance of relying on data from \textit{official }or\textit{ reputable sources}, such as: "scientific literature, different institutions that publish data, UN [United Nations], WHO [World Health Organisation], EU [European Union], World in Data...the more raw the data, the better" (P7).

During the investigation process, fact-checkers evaluate and interrogate the retrieved data to narrow down to documents that meet their high standards of credibility and robustness: 
\begin{quote}
    "You find different sources, and you have to assess whether they're okay or not [...] looking at the data that is contained, looking whether the report, or the dataset to the question you're trying to answer. Looking at the methodology, whether [...] there were some articles on controversies about this report saying that... it's methodologically incorrect [...] assessing whether the database is, in fact, from the correct time-frame you are working on [...] I think you could somehow quantify it and have a concrete measure of the quality ... It's about relevance, whether the source, in fact, is relevant to the thing you're trying to assess. And whether the source is in any way modified [...] the less modified, the more raw, I think it's better." --- P7, Director \& Journalist, Poland
\end{quote}
It is important to note here that in contrast to fact-checker emphasis on primary sources, automated fact-checking techniques tend to rely on secondary sources (e.g., Wikipedia \cite{gad2019exfakt}, news articles \cite{ferreira2016emergent,pomerleau2017fakenews}, or existing fact-checking articles \cite{atanasova2020generating,kotonya2020fcpublichealth}) in part because primary sources require more complex multi-modal processing. As fact-checkers collect evidence, they evaluate each source and compare sources to each other in terms of coverage and quality, building a coherent knowledge base. In contrast, automated tools use stance detection to evaluate evidence documents, considering each document on its own terms.
This contrast represents challenges for the adoption of automated fact-checking. Fact-checkers may be unlikely to use a system that relies on secondary sources, while automated fact-checking systems may struggle to retrieve important primary evidence to reach a verdict.

\subsubsection{Automated tools for evidence retrieval} 
Participants used various automated fact-checking tools for assistance in identifying sources, extracting evidence from large text, video, or audio files, and, in some cases, retrieving evidence.
For example, P4 and P7 discussed using AI tools for geolocation estimation, P1 described a colleague using a tool for facial recognition. P3 and P9 mentioned Bot-o-meter, a tool that identified bot-based Twitter/X accounts. 
Participants (P1, P3, P4, P5, P7, P8) also discussed negative experiences using LLM-based chat-bots such as ChatGPT and Microsoft Copilot for evidence retrieval:

\begin{quote}
    "We've tested it, you know we asked ChatGPT a lot of questions about public figures in the Ukrainian context and it provided us with wrong answers ... shows you that you have to double-check; you cannot just trust that" --- P1, Investigative Journalist, Ukraine
\end{quote}
Instead, ChatGPT was more useful for summarising or sorting through evidence. For example, P7 described creating "custom GPTs" to assist with extracting evidence from large files:
\begin{quote}
    "In [Chat]GPT, you can upload the documents, a file, and you can then converse with it. You can just ask the questions regarding the document"
    --- P7, Director \& Journalist, Poland
\end{quote}

Overall, fact-checkers tended to be wary of relying on automated tools because they needed to be able to check that the evidence is relevant and from reliable sources. 
Perhaps the biggest barrier to tool usage was language-specific limitations of current AI systems.
Fact-checkers who worked with languages other than English (P2, P6, P8, P9, P10) noted that many automated fact-checking tools were less reliable (if at all) for non-English text and video, as well as for non-Western accents in English.
As P9 explained: 
\begin{quote}
    "Most of the AI tools that have been developed so far do not understand African accents. So they were not able to identify [...] the way I talk, I have my Nigerian accent. I am not British and not American; I'm not European either!" --- P9, Investigative Journalist, Nigeria
\end{quote}
P2 explained that while tools trained in English can be ineffective, where tools in local languages are available, they are appreciated:
"There was a tool here in Argentina that was called TranscribeMe... and it was a tool curated by Argentinians. So that worked awesomely."

At the same time, two fact-checkers (P8, P10) also mentioned that the quality of misinformation generated by AI (e.g., deepfake video or audio) was also noticeably lower for languages other than English, making it easier to identify and debunk as false. For the same reason, AI-generated disinformation in these languages can be less prevalent: 
\begin{quote}
"Most of the tools [...] people are using to generate deepfakes [...] they haven't been trained on Indian language models [...]  
I'm not 100\% confident ... [but] that could be why we did not see much deepfakes, and... maybe it's also got to do with the cost as well [...] 
let's say one party wants to defame the other party by creating, you know, generative [AI] then they have to incur a lot of cost." --- P10, Fact-Checker, India
\end{quote}
LLMs are known to perform less well on low-resource languages \cite {bang2023multitask,lai2023chatgptlanguage}, though recent research has demonstrated that their performance can be improved by increasing model size \cite{ahuja2024megaverse}.
However, larger LLMs also tend to require paid subscriptions (e.g., ChatGPT) or, when open source, remain expensive and compute-intensive to run. 
Poorer performance on non-English claims is an acknowledged issue by automated fact-checking research, however progress remains slow \cite{zhang2024needlanguagespecificfactcheckingmodels}.

\subsubsection{Explanation needs for automated evidence retrieval} 
Although evidence retrieval is by no means a low-stakes task, it is a laborious one, so fact-checkers seemed willing to use a number of tools without needing too many explicit explanations. The explanations they required tended to focus on just enough information to verify claims and check reasoning.
Most fact-checkers described having a robust editorial process, in which their work is cross-checked by peers and scrutinised by the editorial team before publication.
In the same way, they needed to verify automated system output by checking that the evidence used by the model is appropriate and that the evidence infers the predicted output:
\begin{quote}
    "If it gave me an output of [...] a claim that it claimed was fact-checked [...] If it gives sources, I would still click through to each of those sources and make sure it says what the AI says it says. But if it's correct, then great, then I can use the content it gave me." --- P4, Investigative Journalist \& Trainer, USA
\end{quote}
In particular, fact-checkers were interested in why a particular document or dataset is seen as relevant, given the claim, or which specific parts of the text or image were deemed important. 

A key challenge for fact-checking is that often, the information required to make a conclusive judgement is not readily available or accessible.
In these situations, fact-checkers contact expert sources, such as officials or academics:
\begin{quote}
    "[If] this is not something I can verify personally, this needs [a] new piece of information. So what you do is you reach out to people who can comment on that. So [...] I reached out to environmental experts and said, is this a realistic number?" --- P1, Investigative Journalist, Ukraine
\end{quote}
This is a much harder challenge for automated systems, which instead must be able to first identify when the evidence required to make a veracity prediction is not available, and then explain what information is missing, so that the fact-checker can seek it out from expert sources.
Previous work has examined the capacity of automated fact-checking systems to identify where there is insufficient evidence to make a veracity prediction  \cite{atanasova2022insufficient}. Incorporating this alongside explanations could be potentially useful for fact-checkers.

\subsection{Verdict decision} \label{findings:verdict_decision}
\subsubsection{Decision-making in veracity prediction}
Everything that fact-checkers do culminates in a verdict. As with any complex decision, despite all the evidence it is often difficult for fact-checkers to articulate how, in the end, the decision is actually made \cite{chen2023intuitionxai,sadler2004intuitive}. Several fact-checkers struggled to articulate their precise reasoning process, referring to gut-feeling or "instinct" (P7). 
Participants described sometimes having an intuitive sense that a claim was false, based on years of experience:
\begin{quote}
    "How do I explain this? Because [...]
    when you've been doing this work for almost 6 years now, your mind just gets ingrained and it processes the misinformation quite quickly." --- P10, Fact-checker, India
\end{quote}
For most, the responsibility of deciding on a final verdict is not an individual one. Fact-checking involves highly collaborative workflows and robust editorial processes \cite{juneja2022human,graves2017anatomy}. P4 recounted "having to do that fact-check process every time we did a new draft of the story [...] I redid the fact-check like 7 times overall".
P3 described his organisation's procedures of peer review and collective decision-making, and their strength in mitigating potential bias in assigning verdicts:
\begin{quote}
    "This [biased decision-making] we try to diminish [...] all people in the team have to agree for that article to be published, because we have different points of view. So when we agree, then there is a bigger chance that there is no risk of bias, because there's always risk of bias." --- P3, Fact-Checker, Poland
\end{quote}
Nevertheless, when pressed, fact-checkers pointed to evidence as the foundation for their decision rationale. 
They described laying out a clear and logical evidence-based argument when communicating the final verdict to the public. Part of the challenge in deciding whether something is true or false, is that most misinformation claims are neither. A good lie, after all, always has a grain of truth in it. 
Fact-checkers (P3 and P7) acknowledged that on one hand, assigning labels such as "true", "false", "half-true" had significant advantages; they are clear and are easily and instantly understood by people.
On the other hand, such labels can gloss over nuance and "make reality much simpler than it is" (P7).
Instead of categorical verdicts, a number of fact-checking organisations (e.g., Pagella Politica;\footnote{\url{https://pagellapolitica.it/}} Full Fact\footnote{\url{https://fullfact.org/}}) publish verdicts one or more sentences in length.
These verdicts have the advantage of being able to specify which parts of the claim are true or false, allowing for more nuanced conclusions.
P7 suggested that more descriptive verdicts may have more powerful long-term effects on readers:
\begin{quote}
    "What does it mean that something is false? [...] I think I would prefer [readers] to remember the verdict in form of text [for example,] `so this claim is against most recent reports.' And it can include words such as `false', `truth', `manipulation', `unverifiable', but [it's] not the sole purpose" --- P7, Director \& Journalist, Poland
\end{quote}
However, P7 also noted that: "text-like claim and verdicts, they will encourage different type of emotional manipulation, meaning, for example, usage of some kind of verbs, which might have not neutral approaches. Which, again, someone might argue that might lead to some kind of polarisation."

Descriptions for verdict labels are essentially explanations -- sometimes, we want them to be simple, while at other times we want more complex answers. The appropriate complexity depends on the particular claim, the context, the audience, and many other factors \cite{Johnson2019SimplicityAccount,Lim2020ExplanatoryMatching}. Claims that require fact-checking can be politically sensitive or controversial, and verdict formats, whether one-word or more elaborate judgements, have simultaneous advantages and drawbacks. Our participants were conscious that even with longer explanations, sometimes all people will remember is the term "true" or "false", although at other times the fuller explanation is what matters, and this is not always predictable.

\subsubsection{Automated tools for veracity prediction} 

When asked about the use of AI tools for veracity prediction, our participants were as sceptical as those in prior studies \cite{juneja2022human,micallef2022true}. They were concerned about bias, lack of nuance, and that the task is simply too complex for an automated system:
\begin{quote}
    "AI can do some technical work. It can maybe verify images or videos or visuals, but it won't be able to explain complex issues like [...] misrepresented [manipulated] quotes, [...] or some wrong accents in a particular statement, [...] you would still need a journalist because you would need that cultural, social, background knowledge and ability to explain things and in a way that's relevant for you and also for your audience." --- P1, Investigative Journalist, Ukraine
\end{quote}
Aware of the potential opportunities and difficulties posed by LLMs \cite{wolfe2024genAIfactchecking}, participants were also concerned about AI hallucination and the reliability of output.
One fact-checker highlighted the epistemological debate about the nature of truth as a key reason they were sceptical about automated fact-checking:
\begin{quote}
    "AI does not really operate on the true, false categories [...] AI is more concerned with just prediction and results... We as humans have been discussing the definition of truth since the ancient Greeks, and I don't really think that such a difficult concept can be applied in code, like truth is something more complex to be reliably coded in a machine." --- P3, Fact-Checker, Poland
\end{quote}
Many claims dealt with by fact-checkers are not straightforwardly true or false, and therefore require detailed and complex evidence retrieval and evaluation. 
This process is extensively documented and an important part of the story that accompanies the verdict. Automated fact-checking systems, in contrast, are primarily concerned with veracity prediction as the ultimate goal, where precision is prioritised over how the verdict was reached. 
One example of this is the phenomenon of shortcut learning in NLP models, in which models make predictions based on dataset biases as opposed to useful features \cite{du2024shortcuts}.

\subsubsection{Using AI tools for detecting AI manipulation}
The introduction of generative AI has resulted in a proliferation of misinformation produced using these tools \cite{dufour2024ammebalargescalesurveydataset}. Our participants felt they needed to use tools to detect such manipulation or at least to "confirm" their own intuitions about AI-generated or manipulated media. P5 and P10 described using tools such as Hive Moderation,\footnote{\url{https://hivemoderation.com/}} DeepFake-O Meter,\footnote{\url{https://zinc.cse.buffalo.edu/ubmdfl/deep-o-meter/landing_page}} and Truemedia.org\footnote{\url{https://www.truemedia.org/}} by uploading media or text to the platforms, which return a verdict on its authenticity, often accompanied by a percent confidence level, such as "Input is likely to contain AI-generated content --- 99.9\%". Some participants described paying particular attention to the level of confidence accompanying an output:

\begin{quote}
    "I think the confidence [...] does help, [...] it makes me a bit more confident as well. I was [...] 50\% sure...that this is likely to be AI-generated. But now the tool is also giving me confidence." --- P10, Fact-Checker, India
\end{quote}
Here too, participants were cognisant that AI tools could be unreliable, especially when the confidence level output did not match their feeling that "something is off" (P5) about a particular piece of media. Multiple participants (P2, P3, P5, P6, P10) described triangulating verdicts by using several AI tools to predict the veracity of the same input, check the same image or video, seek better quality versions of the same video, and other approaches.
Fact-checkers were ultimately aware of being in an "arms-race" between misinformation production and detection, particularly as generative models improve and cues to detecting manipulated media become harder to identify with the human eye:

\begin{quote}
    "The purveyors of deepfake technology... they are also getting better... initially [...] their images used to have those extra fingers, those waxy textures, which has improved a lot with [...] the new versions of Midjourney and Dall-e coming in... they've upped their game, so the detection tool also needs to get better and better." --- P10, Fact-Checker, India
\end{quote}
Most importantly, several participants (P2, P3, P5, P6) pointed to the lack of explainability as a contributing factor to feeling these tools are unreliable. 
P3 in particular highlighted the lack of warrant provided by automated fact-checking tools:

\begin{quote}
    "We do not have reliable epistemology of AI, because it's opaque... it's not explainable, it's not transparent, then we have no reason to trust it unconditionally. We have no warrants to the beliefs that we acquire from our different agents, they do not have a warrant... maybe they have a justification, but they do not have a warrant." --- P3, Fact-Checker, Poland
\end{quote}
P5 described her concerns about the risk of unintentionally misleading her readers as a result of not understanding how the automated tools worked, and pointed out the irony of using uninterpretable AI tools for fact-checking:

\begin{quote}
    "And people... they used to trust things on the Internet, and we show them that they shouldn't. And it's good to have [...] trusted tools, not the black box tools to explain them, that something cannot be trusted [...] It's bizarre that we use those tools, that just because they look fine... I need to trust them, but I would love to give like something better for my readers... They have to trust me, and the tools that I use as well." --- P5, Fact-checker, Poland
\end{quote}
While fact-checkers are increasingly forced to rely on AI tools, especially in tackling misinformation produced through generative AI, they struggle with understanding how these tools work, lacking explanations they find useful and finding the tools unreliable when they do use them. 
While the accuracy of current tools will likely improve, understanding the rationale for predictions and when they should and should not be relied upon remains paramount.

\subsubsection{Explanation needs for automated veracity prediction} \label{veracity_explneeds}
While our participants clearly sought explanations for how automated systems might assign verdicts to misinformation, they often pointed to three aspects that an explanation should include - the why, the how, and the who. 

\textit{The Why: Explaining model verdicts.} 
As well as explaining their process, automated fact-checking models must also explain how the evidence they gathered justifies the predicted verdict.
A common theme was that fact-checking explanations must provide the sources used in verifying the claim (P2, P3, P4, P5, P6, P7, P9, P10).
More specifically, fact-checkers discussed the function of explanations as cues for where they needed to look and check for themselves; and providing citations and evidence that they can verify.

They were particularly interested in local explanations referring to specific elements of the evidence that lead to the model's verdict, such as sections of audio that indicate manipulation, or frames in video where generative-AI artefacts were detected. 
Such explanations were useful not only for identifying rationales for the tool's output, but also for expediting fact-checking:

\begin{quote}
    "If... something's going on in the audio, like you can represent audio in this waveform, right? And you can tell if it was edited [...] 
    And if you're talking about the video this tool can show me the certain frames where you see that there's something going on with the lips of the person who's appearing on the video, or the movements are not right, but it's really specific. 
    [...]
    That would give me also like a picture [...] that's something that I can show my readers like [...] you should see this little one-second clip really closely because it shows the flaw of the AI tool... that will make my job easier cause I don't always have the time and expertise to look so closely at the media." --- P5, Fact-Checker, Poland
\end{quote}
Although there are automated methods that can highlight regions where content is contributing to a prediction, they are limited to known behaviours and interactions, and overall remain in their infancy for video \cite{cizmeciler2022leveraging}.

While measures of certainty that models output alongside a verdict were important, several participants (P2, P3, P5, P7) did not understand how such scores were calculated, and how they should be interpreted, "What does 65 versus 74 confidence mean?" (P7).
One fact-checker described how quantification of model uncertainty alone raised more questions than it answered, and imagined how verdict uncertainty could be conveyed in a human-like way, such as highlighting issues with the reliability of sources used in deciding the verdict:

\begin{quote}
    "If [an automated fact-checking tool] said to me, `I'm only 70\% sure because...' [...] for example, `here is a report that mentions data, you need, but this is a report done by a scientist who has often had his papers retracted.' If it gave out this explanation of the uncertainty, it would be amazing... But just a grade of certainty, like this is 70\% good data, no, because I would be more sceptical, not less." --- P3, Fact-Checker, Poland
\end{quote} 

\textit{The How: Explaining model processes.} 
As we have noted, fact-checkers consider the path to the fact-checking verdict as important as the verdict itself. It is perhaps unsurprising that our participants were keen to understand the process of how an automated fact-checking tool arrived at the verdict.
In the absence of explanations, some developed their own intuitions or folk theories about how automated tools worked, such as certain words or phrases that were frequently picked up. 
However, some fact-checkers (P2, P4, P5, P6) expressed concerns that explanations of AI processes would be too complex or time-consuming to understand.
For example, although P2 thought it would be "useful to know, both processes, to know everything", she could also "imagine not understanding a lot of the process, if it's too technical."
P4 recalled one automated tool that she used, which had included a link to an article which described the technical approach used:
\begin{quote}
    "To their credit, they explained it [automated claim-detection tool], but it's like a 20-page academic paper about deep learning... that's great, that they have that kind of transparency [...] I'm relatively technically competent among journalists... I have no idea what that means. So I think it's great for transparency, but realistically, no one's gonna understand that, at least not journalists." --- P4, Investigative Reporter \& Trainer, USA
\end{quote}
Instead, our participants sought explanations of automated systems expressed in terms of human fact-checking processes. 
For example, wanting to know all the sources and pieces of evidence that were "checked" by the model, how they were checked, and a rationale for how evidence was selected:
\begin{quote}
    "If it did give me some specifics like, `I used this tool to search all tweets they ever posted in the last 12 months. And then I narrowed it down to these keywords' [...] that's something the journalist can understand, and could be really helpful for them to know ... specifically, how did it do that? Cause... the journalist knows what tweets are." --- P4, Investigative Reporter \& Trainer, USA
\end{quote}
P4 was curious about how the tools and the models were designed but did not see this as relevant to fact-checking itself. Instead, fact-checkers wanted explanations that would align with the fact-checking process, showing the work "like we do as human beings" (P8).
While such an explanation format may seem straightforward, \textcolor{black}{it would be challenging to produce from a technical point of view, as the language models that underlie automated fact-checking systems do not employ human-like logical reasoning.}
Yet fact-checks require transparency, and thus, showing the process is paramount for fact-checking tools to be useful.

\textit{The Who: Explaining model training.} 
Seven participants (P1, P2, P4, P5, P7, P9, P10) expressed that knowing about the origins of an automated fact-checking system, and in particular, knowing the data that the system was trained on, was essential in knowing when to rely on the model. Fact-checkers are acutely aware of their own biases and take steps to mitigate these. The same is true of the tools they might use --- they need to understand the potential for biases the tools may have so they can mitigate these too.
Knowing the sources of training data, the amount of training data, and the languages contained in the training data was seen as critical to identifying information gaps and calibrating when to rely on the model.
The latter point was strongly evident for participants who worked in non-English speaking countries, and also for participants from Africa who worked with English, but found that automated tools were ineffective in picking up African accents and region-specific nuances. P9 expressed his hope that "organisations working towards building AI tools [...] could consider using more African datasets in developing their tools."

Increased model transparency was also important from an ethical perspective: \begin{quote}
    "I would like to understand who created this particular application or software? Where did they get all the data? How they were testing it and so on [...] So founders, finances, data sources, the algorithms for generating different responses." --- P1, Investigative Journalist, Ukraine
\end{quote}
The demand for disclosure of training data is an ongoing debate in the AI community, with proposals such as data sheets for datasets \cite{gebru2021datasheets} and model cards \cite{mitchell2019modelcards}, creating structured ways such information could be disclosed. 
Yet there remains a challenge of not only demanding transparency from the providers of many of the available tools, but also with bridging the gap between the technical idea of what is important to disclose with respect to models and data and what fact-checkers expect given their own professional context.

\subsection{Communication: explaining fact-checks} \label{findings:communication}

\subsubsection{Decision-making in explaining and communicating fact-checks}
After reaching a verdict, fact-checkers typically write an article consisting of the claim, the verdict, and the explanation of how they arrived at the verdict. Participants related how explanations are key to communicating their work to the public:
\begin{quote}
    "Without [explanations] it would be censorship. In my opinion, if you have a tool which would be just assessing true-false, true-false without explaining... this would not be okay [...] if a big company would employ a tool that would be just a classifier that would [be] just striking, and I think some of them already do, I don't think it's okay. I think there needs to be even [a] simple explanation why the decision was made" -- P7, Director \& Journalist, Poland
\end{quote}
As such, the same considerations fact-checkers have for evaluating their own explanations of the final verdict are relevant to how we might need to think about the explanations provided by automated fact-checking systems. There are three main considerations that our participants brought up in how they construct explanations of the fact-check for the public: replicability, the type of verdict, and the complexity of the claim fact-checked. 

\textit{Explanations must be replicable.}
Replicability was identified as the most prominent theme. All participants discussed the importance of including links to sources, public data, and tools used or referenced in the fact-checking process, with the view to including sufficient information that readers can reconstruct and replicate the fact-check.
P1 described this practice of providing and signposting evidence as "self-explanatory", while P6 identified this commitment to "show the work" as the central tenet of fact-checking: 
\begin{quote}
    "I think fact-checking is a different branch [of journalism] for that reason... Our work has to be shown, like the kind of math class thing back in the day, you know. Show your work. How did you arrive at this? We have to be able to do this." --- P6, Senior Fact-Checker \& Project Manager, Ireland/USA
\end{quote}
To date, explainable automated fact-checking techniques have focused on explaining only the predicted verdict, that is, how the evidence proves whether a claim is true or false \cite{kotonya2020explainablesurvey}. Yet focusing on replicability goes beyond providing a set of reasons for why a judgement was made. The tools and the relevant evidence should enable replication of the process: 
\begin{quote}
    "When we publish something, we want our readers to [...] feel that they can trace every step that we took. Like, go on this website, click, this thing, and now we use this database, search for page 43. Now you enter this website, compare this [...] we spell out every step that we took to have completely replicable analysis. So when we use external tools in our analysis, we also want people to be able to enter this website and see for themselves why we used it, and what judgement it spelled out, and why it convinced us, so why should it convince them." --- P3, Fact-Checker, Poland
\end{quote}
This means that automated explanations must be able to communicate the sources of evidence (e.g., databases, webpages) that were checked, how they were checked (e.g., search terms used), why sources were selected as evidence (e.g., source reliability or relevance), to the extent that it is possible to reproduce the fact-check.

\textit{Explanations are verdict-dependent.}
The category of the verdict assigned to a claim influenced the form of explanations participants provided to readers. The common binary of either completely true or completely false was the most straightforward to explain. 
The typical first step in their investigation involved attempting to verify a claim in good faith, and, if unable to do so, first explain why the claim is false, where the false information came from (e.g., whether it materialised from a misquotation, cherry-picked data, or pure fabrication), before turning to an explanation of why the correct information is true.
Claims that were half-true, or contain some element of truth, tended to be more difficult to fact-check and explain than claims that were completely false, as they tended to require additional context and more nuanced analysis, as well as longer and more complex explanations.
\begin{quote}
    "If something is false, we're gonna spend the article explaining how we know this is false... [for] misleading... the underlying logic would basically be, this doesn't prove what you think... there's missing parts here... you might want to consider" --- P6, Senior Fact-Checker \& Project Manager, Ireland/USA 
\end{quote}
Another challenging verdict was the inconclusive "unverifiable" or "unproven" claim. 
Some fact-checkers (P5, P8, P10) recounted being forced to abandon claims due to a lack of available information.
Others felt it was still worthwhile to publish such verdicts because "the lack of information is also information. It also explains to our reader that maybe some commonly-repeated myth is not actually based on good data" (P3).
Similarly to the emphasis placed on replicability, fact-checkers found it valuable to transparently lay out all of the existing evidence, allowing the readers to draw their own conclusions.
\begin{quote}
    "in such cases we put all the evidence, all the facts out there. So it's left for the readers to now decide, `Okay, these are all the facts.' Then you can make up your mind or opinion about it." --- P9, Investigative Journalist, Nigeria
\end{quote}
Current technical approaches to explainability tend to present the same type of explanation, no matter the context, and this clearly needs consideration.

\textit{Explaining complex stories.}
Complex verdicts require complex stories, and complex stories are challenging to tell. All participants commented that it is not possible to predict a reader's background knowledge and tailor the article accordingly. As a result, they have to assume minimal audience background knowledge and provide as much detail as possible.
For example, claims related to economic and legal issues were highlighted as particularly challenging to explain in an accessible manner, due to their complexity.
Scientific and health-related claims were also difficult to explain in a convincing way, partly related to their complexity, and partly "because there's just a lot of people who are hardcore conspiracy theorists.
So [...] regardless of data that you provide to them, then it's just really difficult to prove anything" (P1). P1 also noted that claims related to history were easier to communicate "because it's more of a narrative format".

Here, emphasis on replicability and clarity of structure were key, such as providing a logically coherent summary at the beginning to highlight the core argument:
\begin{quote}
    "This summary needs to be a little bit like logical reasoning in philosophy [...] Argument 1, Argument 2, Argument 3, they need to completely justify the last point, that is the conclusion, and in the conclusion, in the last bullet point should be the grade: Is it true? Is it false? Is it manipulation? Is it unverifiable? So the arguments in the bullet point list should lead to the conclusion, and then, if it doesn't, then we stop and work on the article more." --- P3, Fact-Checker, Poland
\end{quote}
Especially where claims and stories were complex, participants acknowledged that despite emphasis on replicability, the explanation they constructed did not reflect all of the research steps they might have performed. Research is often not a straightforward enterprise. The explanation, however, is a straight path through what can often be a labyrinthine process: 
\begin{quote}
  "Taking care that the story and article is written in a very logical way... research can be chaotic [...] making sure that the story will be step by step." --- P7, Director \& Journalist, Poland.   
\end{quote}
Thus the more complex the claim and the verdict, the more our participants emphasised structure and clarity of explanation. Intimately aware of the challenge and sensitivity of their task, they both sought to provide transparency and to ensure that as many as possible in their audience could easily grasp the point. This consideration of an unknown audience is key to the rhetorical purpose of explanations, but rarely directly considered when these are created automatically. 

\subsubsection{Automated tools for communicating fact-checks}

The availability of LLM-based tools that can produce well-appointed text has led to much debate on the future of journalism in general \cite{pavlik2023collaborating}. Fact-checkers were also acutely aware of these systems and their capabilities. Participants for whom English was not their first language (P2, P3, P5, P7, P8) noted that they at times used LLM-based chatbots such as ChatGPT and Microsoft Copilot\footnote{\url{https://copilot.microsoft.com/}} to improve the quality of their writing, summarise fact-checks (e.g., bullet points or a short paragraph at the top of the article), edit articles, and disseminate fact-checks.

For example, P7 used the paid subscription to ChatGPT and mentioned using the tool for "cross-checking, whether there is something to be improved [...] sources would have to be verified, but still... the argumentation could be relatively good". P8 mentioned using ChatGPT to write scripts for video versions of the fact-check article to be disseminated on social media.

Participants (P4, P5, P7) also mentioned ethical concerns about using AI in fact-checking, specifically, the use of news articles for LLM training without permission from journalists:
\begin{quote}
    "In the journalism world, people are relatively sceptical in terms of ChatGPT, like generative AI, because, it's stolen work, the journalists' work, to learn, without compensating them, and that's the reality." --- P7, Director \& Journalist, Poland
\end{quote}
Such ethical considerations could also manifest as stigma for those using the tools: 
\begin{quote}
    "There's a sense of pride in not using these kinds of tools [...] I remember we had some discussion about ChatGPT, and [...] people were like, `No, don't use it because somebody will tell us, you're using... some other people's work.' --- P7, Director \& Journalist, Poland
\end{quote}
The fact that LLM-driven systems are only able to produce text because people have produced a lot of this text first is intractable. Where these tools could be useful, such considerations remained important, especially in an environment that constantly deals with sensitive and political issues while often struggling with precarity (see also \cite{wolfe2024genAIfactchecking}).

\subsubsection{Explanation needs for fact-checkers}
Our participants made it clear that there is a difference between the kinds of explanations they might furnish for their readers and the explanations that they themselves required of the systems they used. 
Fact-checkers require a high threshold of certainty in their verdicts to maintain the confidence of their readers and the general public in their work \cite{micallef2022true}. 
While our participants attended to measures of model uncertainty provided by automated tools, 
for several (P3, P5, P7) numerical measures of confidence in the form of percentages or scores were unhelpful and disconnected from how fact-checkers reason about the reliability of evidence. Measures of confidence are not the same as measures of quality of the evidence used:
\begin{quote}
    "Confidence, depending how, if you would have an idea how to present it, would be good, maybe showing the thought process of a tool [...] some kind of idea for showing the quality of the research that the tool has done." --- P7, Director \& Journalist, Poland
\end{quote}
In a similar way to how fact-checkers have their work cross-checked and scrutinised by peers and the editorial team prior to publication, our participants expected automated tools to provide information so that they could verify model decisions by checking that the evidence used by the model is appropriate and infers the predicted output.
To facilitate this, local explanations of model outputs that refer to specific elements of evidence, such as excerpts of documents or sections of images that contributed to the prediction would be helpful:

\begin{quote}
    "I would love to have, for example, timestamps, where [...] I have the video, and this AI recognition tool can give me, for example, timestamps and screenshots of this video in those times... something like, `Hey, at this moment you can see...', and you can also show your readers that there is some evidence of AI manipulation." --- P5, Fact-checker, Poland
\end{quote}
Focusing on specific, checkable parts of evidence may also alleviate concerns about automated tools being too complex to explain to journalists with limited technical backgrounds or expertise, by explaining model decisions in fact-checkers' own terms.

Finally, faithfulness is an important concern when generating automated explanations. Yet there is a significant difference between how automated systems arrive at veracity prediction and how fact-checkers make verdict decisions. Our participants had mixed views on how complete and faithful a model's explanation should be to the underlying decision-making process.
Contrasting opinions highlighted practical tensions between explaining automated fact-checking in a manner that reflects fact-checker processes versus faithfulness to the model's own decision-making process.
Some participants (P5, P6, P8) expressed a desire for explanations that would be easy to understand and explain to readers, even if they were not representative of the tool's inner workings:
\begin{quote}
     "I understand... maybe it's not working like that, inside this black box. But that would be really useful [...] like to show people what are the clues, that give you this high chance that this video is manipulated." --- P5, Fact-Checker, Poland
\end{quote}
Others worried that if the gap between the explanation and how the model actually works was too large, then the explanation may not be reliable:
\begin{quote}
   "I would worry like well, is it hallucinating the methodology [...] I would have to be pretty confident in the tool to use it that way." --- P4, Investigative Reporter \& Trainer, USA
\end{quote}
This mix of opinions reflects ongoing debates in the field of explainable AI about what constitutes a good explanation \cite{jacovi-goldberg-2020-towards, liao2022humancenteredexplainableaixai}. Our data demonstrate the need to accommodate different audience needs for explanations and the fact that these can change with context, where both model faithfulness and understandability must be considered. 

\section{Discussion} \label{discussion}
For each stage in the fact-checking process, our findings revealed key criteria for how fact-checkers make decisions (\textbf{RQ1}), how they use automated tools (\textbf{RQ2}), and the information that explanations of these tools should contain to support their work practices (\textbf{RQ3}).
In this section, we consider the current findings with regard to existing work in automated fact-checking, identify their implications for technical and HCI research and opportunities to help bridge existing gaps. Finally, we suggest future work in automated fact-checking and reflect on the broader implications of AI in fact-checking, as the system of human fact-checking continues to buckle under the increasing onslaught of misinformation. 

\subsection{Fact-checker explanation needs versus automated fact-checking capabilities}
The process of fact-checking and communicating verdicts is complex and increasingly requires a wide range of AI tools, partly due to the sheer volume of misinformation and partly because much of this misinformation is produced by the same tools used for its detection. 
Ironically, the capacity to produce misinformation continues to outstrip the capacity to detect it. Moreover, veracity detection tools are insufficient alone; they require explanation and automatic explanation generation is more technically limited still. Our participants explained the complex \textcolor{black}{judgments that impacted their use of automated tools} and the shifting explanation needs during the different stages of the fact-checking process.
 
In line with prior work \cite{micallef2022true,juneja2022human}, fact-checkers in our study noted the time-consuming nature of identifying and triaging claims. Currently, claim detection is perhaps the task that is best equipped with existing tools \textcolor{black}{(see \cite{Das2023state} for a review). This is a relatively low-stakes task, requiring fewer and less detailed explanations of the underlying processes}. However, rationales for how and why particular claims were selected can enable fact-checkers to identify flaws or biases in tools. 
Existing explainability techniques such as importance rankings for claims and feature importance scores \cite{ribeiro2016LIME} are easily implemented for this task.

Stakes increase for evidence retrieval, where \textcolor{black}{fact-checkers described the central role of evaluating the relevance and credibility of evidence and the deficiencies of automated tools for this task.} Useful explanations could allow fact-checkers to swiftly verify whether a given piece of evidence is reliable and\textcolor{black}{, that it} has been correctly interpreted by the model. 
While existing techniques such as attention highlights for text \cite{popat2018declare} and saliency maps for images and videos \cite{petsiuk2021saliencymaps} are sometimes used for these purposes \cite{lim2023xai}, there are unresolved issues with regard to their faithfulness to the underlying model and legitimacy as explanations \cite{jain2019attention,bibal2022attentionexpl,adebayo2018sanity}. There are opportunities here for design to present output from existing tools in more useful ways, as HCI research in this area is limited.

Reflecting the intricate judgments and decision-making processes described by fact-checkers, veracity prediction is the most complex task in automated fact-checking, and explanations for these decisions require more detail and complexity.
Fact-checkers seek specific, local explanations that highlight key parts of the evidence documents and link them to the verdict. Current approaches, such as feature importance scores \cite{ribeiro2016LIME,lundberg2017SHAP} or summaries of evidence, may be able to at least partially address these needs \cite{atanasova2020generating,kotonya2020fcpublichealth}, however, issues have been raised about their stability \cite{alvarezmils2018robustness} and reliability \cite{slack2020fooling}. 

Current approaches to communicating uncertainty (i.e., simple percentages) also seem insufficient as the fact-checkers found it difficult to relate them to the quality of evidence and source reliability and to compare across tools. 
Although prior work has explored human-interpretable confidence scores using case-based and counterfactual explanations for classification tasks \cite{van2020interpretable,le2023explaining}, 
reliable, human-like explanations of uncertainty represent a challenge. For example, while LLM expressions of uncertainty can reduce people's over-reliance on them \cite{kim2024LLMuncertainty}, LLMs tend to overestimate their own confidence \textcolor{black}{\cite{tanneru2024uncertaintyLLMs,steyvers2024calibrationgapmodelhuman}, so these scores are often unreliable.}
Explanations must also reflect the context, nuance, and complexity in verdicts and claims themselves.
We see this in the distinction between static, temporal, and dynamic facts in LLMs, which may impact the desired level of detail in an explanation
\cite{marjanović2024internalconflictcontextualadaptation}.

The most significant challenge we identified was the fact-checker desire to align model processes with their own approaches, though the logics are fundamentally different. 
Part of the reason for this requirement is that fact-checking is not just about the verdict itself, it is also about offering a replicable process for reaching it.
This requires explaining multiple steps in the fact-checking pipeline (evidence retrieval and veracity prediction). While there is existing work on multi-hop \cite{Ostrowski2021p536} and chain-of-thought \cite{pan-etal-2023-fact} reasoning, these continue to fall short of the complex logic required to explain real-world claims \cite{Das2023state,kotonya2020explainablesurvey,aly2023qanatver}.

Our findings extend evidence of how fact-checkers utilise technology in their work (see \cite{juneja2022human,micallef2022true,wolfe2024genAIfactchecking}) given the recent rise of generative AI and suggest that previously observed resource, financial, and linguistic inequities may be exacerbated by the prevalence of LLMs.
While prior research has found that fact-checker use of tools is "fragmented" \cite{micallef2022true}, our findings suggest that this may be partly due to the lack of explanatory information these tools provide, leading fact-checkers to hedge their bets by using multiple tools for the same purpose.
Many of our participants expressed similar scepticism about automated fact-checking as observed by previous studies \cite{micallef2022true,juneja2022human}, however, we also noted some optimism, particularly in terms of AI's potential to reduce workloads.
Finally, our findings provide some limited support for previously-proposed criteria for explainable fact-checking \cite{kotonya2020explainablesurvey,jacovi-goldberg-2020-towards}.
Our participants expected explanations of fact-checks to mirror human processes. That is, explanations should be \textit{coherent} (follow natural laws) and \textit{causal} (follow logical reasoning), as well as \textit{context-full}, that is, presented alongside the corresponding claim and evidence \cite{kotonya2020explainablesurvey}.
Fact-checker concerns about bias in automated fact-checking also supported the proposal that \textit{impartiality} is an important consideration for fact-checking explanations \cite{kotonya2020explainablesurvey}.
The property of parsimony in explanation was reflected in the need fact-checkers expressed for specific explanations that are easily verifiable. However, the emphasis that fact-checkers placed on verifying and replicating the model's processes suggests that explanations with these necessary details should not be sacrificed for brevity. As such, key human-centred requirements for explanation of automated fact-checking systems must be \textit{replicable}, \textit{verifiable}, and able to explain \textit{uncertainty and information gaps}.

\subsection{Design implications: Connecting automated fact-checking tools to fact-checker processes}
Our findings highlight fundamental differences in the goals and perspectives of fact-checkers relative to automated fact-checking approaches. 
\textcolor{black}{Based on these disconnects, we make the following recommendations for the designers of automated fact-checking technologies for fact-checkers.}

\subsubsection{The importance of primary sources}
The most obvious disconnect between manual and automated fact-checking is in evidence assessment. 
Fact-checkers stressed the significance of primary evidence like quotes, official records, datasets, photographs and video and audio recordings.
In contrast, current automated fact-checking approaches and machine learning (ML) methods more broadly to retrieve, process and reason about unprocessed, unannotated multi-modal data \cite{akhtar2023multimodal} tend to rely on secondary sources such as news articles \cite{ferreira2016emergent,pomerleau2017fakenews}, existing fact-checking articles \cite{alhindi2018evidence,atanasova2020generating,kotonya2020fcpublichealth}, Wikipedia \cite{jiang2020hover,schuster2021factcontrastive}, or heterogenous online sources \cite{augenstein2019multifc,schlichtkrull2024Averitec}.
Fact-checkers cautioned against using such secondary sources as they are susceptible to bias or flawed methodological reasoning, and may have limited direct relevance to the claim. 
Secondary sources may be easier to use, as they incorporate information retrieval methods which rank sources by popularity, but they entail inherently limited and biased data from the fact-checker point of view. Evaluating secondary sources requires cultural understanding to be able to read between the lines, while primary evidence is, of course, messy, unstructured, and potentially missing or incomplete. 

At a minimum, automated fact-checking must evaluate evidence quality, weigh relevance and reliability as fact-checkers do, and explain these assessments alongside the verdict.
Previous (pre-LLM) work has examined classifying media bias, factuality, or credibility of news sources \cite{baly2018factualitybias,baly2020detectbias,baly2020written,yuan2020credibility},
while more recent work has focused on providing more in depth assessments of source reliability \cite{schlichtkrull2024generatingmediabackgroundchecks}. To our knowledge, such efforts have yet to be integrated into explainable automated fact-checking tools. \textcolor{black}{These techniques also fail to sufficiently address the complexity and sophistication that fact-checkers are seeking.}

\subsubsection{Beyond binary verdicts}
Although publishing categorical verdicts (e.g., "true", "false", "misleading", "unverifiable", etc.) remains a convention alongside fact-checking articles, such labels are reductive, particularly given that misleading claims are rarely straightforwardly false. 
Some fact-checking organisations publish verdicts of one or multiple sentences, which can address nuance in complex claims.
While this "less confrontational" approach could increase the effectiveness and informativeness of fact-checking,\footnote{see Full Fact's (UK-based fact-checking organisation) justification for this approach: \url{https://fullfact.org/about/frequently-asked-questions/\#ratings}} it contrasts with the typical approach for automated fact-checking (and ML classification tasks generally) of labelling inputs according to clearly-defined categories.
In fact, for the sake of simplicity, data from fact-checking organisations that do not publish categorical verdict labels are often excluded from automated fact-checking datasets \cite{gupta2021xfactmultilingual,augenstein2019multifc}, and the multi-label categorisation systems used by fact-checking organisations remain a challenge for automated fact-checking techniques, often "solved" by collapsing multiple labels (e.g., mapping the verdicts "mixture", "unproven" and "undetermined" onto the label "not enough information" \cite{hanselowski2019corpus,kotonya2020fcpublichealth,gupta2021xfactmultilingual}; see \cite{guo2022survey} for a review).

\subsubsection{Show the work}
These differences in approaches are important because for fact-checkers, communicating the pathway to the verdict is just as important as the verdict itself. 
Fact-checkers emphasised that documenting the steps and inferences made at each point in the fact-checking process was central to explaining how they reached their final conclusions. In other words, explaining the pathway to the verdict was seen as paramount for maintaining credibility.
Explainable automated fact-checking continues to prioritise the verdict through veracity prediction, focusing on highlighting important tokens \cite{popat2018declare}, or providing rule-based \cite{ahmadi2019explainable} or text-based summaries \cite{atanasova2020generating,kotonya2020fcpublichealth} of how the evidence justifies the verdict.
\textcolor{black}{Automated fact-checking systems are not required to be oracles, but should attempt to explain their processes to be verifiable and useful to fact-checkers.}

\subsection{Future directions for human-centred automated fact-checking}
The significant disconnects between the current state of automated fact-checking and the practical realities faced by fact-checkers clearly call for more interdisciplinary and critical efforts to reconsider the goals of such research and the broader societal implications of AI's increasing role in fact-checking and journalism.

\subsubsection{Future work in explainable automated fact-checking}
Our findings suggest several promising avenues for future work in explainable automated fact-checking for each step in the fact-checking pipeline.
For time-intensive tasks such as claim detection and evidence retrieval, in which volume and velocity are main concerns for fact-checkers, there are unexplored opportunities for HCI research to examine how explanations for AI systems (e.g., for claim filtering, information extraction, source evaluation) can better support fact-checker ability to determine the credibility and reliability of their recommendations and how such explanations could be designed and delivered more effectively.

Successful adoption of automated fact-checking systems for veracity prediction hinges on how well they integrate into fact-checker work processes.
A key challenge is that current automated fact-checking systems simply do not follow human logic. While it is to some degree possible to force them to do so, for example using neuro-symbolic approaches, this has only been successful in highly controlled settings with artificially created datasets \cite{aly2023qanatver}.
Understanding how to bridge machine logic and fact-checker practices requires stronger human-centred approaches, to develop automated fact-checking systems that better align with fact-checker reasoning.
Systems must also provide explanations with practical utility for fact-checkers. For example, although people are sensitive to LLMs indicating some degree of uncertainty in their outputs (e.g., "I'm not sure, but...")  \cite{kim2024LLMuncertainty,steyvers2024calibrationgapmodelhuman,tanneru2024uncertaintyLLMs}, generating explanations that address the sources of uncertainty, and how it may be resolved remains unexplored. 
While technical approaches can compute uncertainty, HCI research is essential to explore how to present it in ways that are useful to a range of audiences.  
HCI research is well-versed in the idea that knowledge is highly situated and context-dependent, especially for fact-checkers and their work \cite{liu2023humancenteredNLP}. While there is some existing work on designing explainable AI systems sensitive to people's contexts, background knowledge and expertise (e.g., \cite{kim2022designing,ehsan2024thewho}), this research is at a nascent stage and often remains far removed from the realities of fact-checking. 

\subsubsection{Ethical considerations and broader implications of automated fact-checking}
Effective automated fact-checking approaches have the potential to address the accelerating scale and speed of misinformation spread. However, dependence on algorithmic systems by fact-checking organisations and journalists may increase scepticism and erode public trust in media institutions, due to flawed or biased systems or perceptions that these institutions are abdicating their responsibilities. For example, reliance on such models raises the risk of neglecting cultural and regional nuances, given the well-documented Western-centric biases in LLMs \cite{cao-etal-2022-theory,cao-etal-2023-assessing,pawar2024surveyculturalawarenesslanguage}. Over-reliance on LLMs for fact-checking may also depreciate the influence and value of journalists' work; recent work suggests retrieval-augmented LLMs tend to favour LLM-generated contexts compared to retrieved (likelier human-written) content \cite{tan2024blinded}.

Similar to other algorithmic systems deployed in complex socio-technical contexts, such as filtering and content moderation, these tools continue to represent the threat of bias and reproduction of existing structural inequalities, increasing the potential of silencing minority voices \cite{dias2021fighting, lee2024people}. While technologies may be developed with good intentions, a lack of grounding in local, contextually situated practices can have a variety of unintended consequences. This is especially true in fact-checking work where claim verification is often not a case of binary true or false, but a nuanced judgement of partial veracity. While professional fact-checkers worldwide appear to have surprisingly similar approaches and processes for claim verification, local knowledge and communal practices are key to doing so successfully \cite{sultana2021dissemination}. Yet, current development of automated fact-checking systems continues to rely more on researcher imaginaries of how fact-checking is performed rather than the actual conditions of fact-checker practice. Given the centrality of fact-checking as a bulwark against disinformation campaigns and the clear need for effective support tools, human-centred approaches to the development of these tools \cite{liu2023humancenteredNLP,wolfe2024genAIfactchecking} are key to facilitating ongoing efforts towards transparency and explainability. 

\section{Conclusion} \label{conclusion}

Our work illustrates that there are critical rifts between the capabilities of current automated fact-checking tools and the goals and needs of fact-checkers.
By outlining the specific information fact-checkers require of these tools at each step of the process, we demonstrate how these tools can be made more useful.
Automated fact-checking systems hold great promise to assist fact-checkers in managing the intensifying volume of misinformation proliferated in online media.
However, without meaningful engagement with fact-checker practices and more interdisciplinary research efforts these tools will not fulfil their potential to support and empower fact-checkers in their work.
For these tools to be truly effective, it is essential that the design of these tools is shaped by a thorough understanding of fact-checker decision processes and explanation needs.
In light of how fact-checking as a practice continues to be targeted for ideological and political reasons \cite{MetaFactChecking2025,PoynterMeta2025}, we also highlight that while fully automated fact-checking may appear as a good solution to misinformation, such systems may be easier to subvert as political winds shift.
Rather than seeking to develop fully automated fact-checking systems, we call for systems that can support and empower fact-checkers in their critical role in society.

\begin{acks}

The authors wish to thank the fact-checkers and journalists who generously volunteered to participate in interviews and without whom this research would not have been possible.

\noindent $\begin{array}{l}\includegraphics[width=1cm]{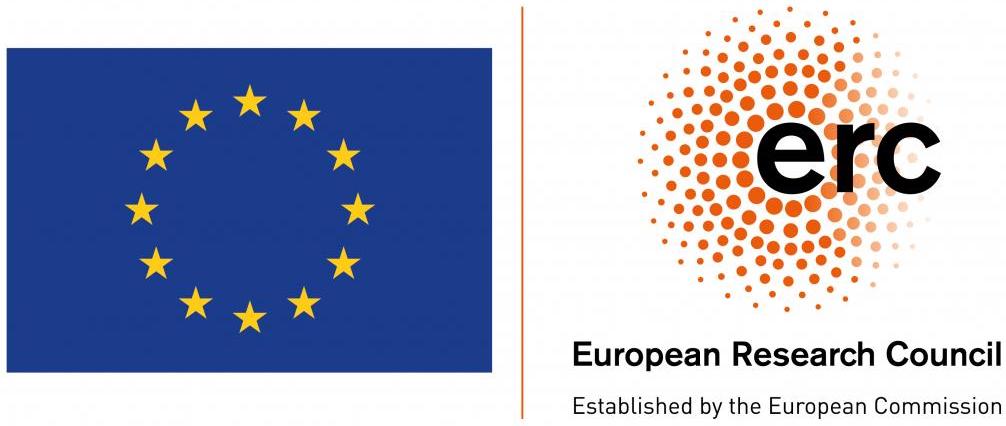} \end{array}$ This research was co-funded by the European Union (ERC, ExplainYourself, 101077481), and supported by the Pioneer Centre for AI, DNRF grant number P1. Views and opinions expressed are those of the authors only and do not necessarily reflect those of the European Union or the European Research Council. Neither the European Union nor the granting authority can be held responsible for them. 
\end{acks}

\bibliographystyle{ACM-Reference-Format}
\bibliography{2_references}

\appendix

\section{Pre-Interview Questionnaire} \label{app:pre-int}

\begin{enumerate}
    \item What organisation (if any) do you work for?
    \item What is your specific role (if applicable)?
    \item How long have you worked as a fact-checker? (years/months)
    \item How many years have you worked for your current organisation (if applicable)? (years/months)
    \item How many years have you worked in your current role?
    \item Do you (or have you ever) used artificial intelligence (AI) tools as part of your fact-checking work? What tools have you used?
    \item What country do you currently work in?
    \item What is your gender?
\end{enumerate}

\section{Interview questions} \label{app:int_questions}
1. Introductory questions
\begin{enumerate}
    \item Could you describe your job and the work that you do?

\item Can you think of an example of a claim that you checked recently? Can you describe the process and steps involved? Backup prompts: What specific goals do you have at each step of the process?
\item How challenging is it to fact-check claims? What is involved? Backup prompt: are there any factors that make some claims particularly challenging to fact-check?

\end{enumerate}

\noindent2. Evidence \& verification

\begin{enumerate}[resume]
\item Using the same example (or a different one), can you describe how you went about searching for and selecting evidence to fact-check that claim? 
\item Now, using the same (or different) example, can you talk about how you decided which veracity label to assign the claim? 
\item What (if any) technological tools do you use to help you with verifying a claim? What are their advantages and disadvantages?
\end{enumerate}

\noindent3. Explanations \& transparency
\begin{enumerate}[resume]
\item Considering the claim you described (or another example), how would you explain how it was given that label? 
\item Were there any challenging aspects involved in explaining the decision?
\item Are there any claim-specific factors that might influence the information you present to people? 
\item Does the audience influence the kind of information you present to people? For example, if the target audience is knowledgeable, or knows very little about the topic. Are there differences in how you would explain a fact-checking decision to an editor compared to a reader?
\item What does transparency mean in terms of fact-checking? How do you (or fact-checkers in your organisation in general) achieve transparency? 
\item What (if any) technological tools do you use to help you with transparency? Are there any tools that you use to produce explanations of the decisions? What are their advantages and disadvantages?
\end{enumerate}

\noindent4. Explanations \& automated fact-checking
\begin{enumerate}[resume]

\item Have you ever heard of any AI-support tools that automate parts of the fact-checking process? Have you ever used any? What do they do, how are they used? What are their advantages and disadvantages?
\item Have you thought about how AI-support tools could be useful to fact-checkers? What steps in the fact-checking process do you think could be supported? How do you think they could be supported?
\item Follow-up question for each step of the process they identify:
Imagine you were using an AI assistant or support tool to assist you with [fact-checking step]. What might it do? What sort of information would you want or expect it to provide? What sort of information would help you to ensure that it is working correctly? 
\item How useful would it be for an explanation of an AI system to describe how it works in general? 
\item What do you think the criteria for a good or useful explainable fact-checking system should be? 
\end{enumerate}

\noindent5. Closing questions
\begin{enumerate}[resume]
\item Is there anything that we haven’t discussed that you think is important or relevant?
\item Is there anything else you'd like to share?
\end{enumerate}

\section{Example themes and codes}
\label{app:themes_codes}
\begin{table*}
    \centering
    \begin{tabular}{l l l l}
    \hline
        \textbf{RQs} & \textbf{Themes} & \textbf{Sub-themes} & \textbf{Example codes} \\
        \hline
        \textbf{RQ1:}  & Evidence quality (\textbf{10}) & Searching for evidence, Contextual clues & `primary sources', `cross-checking' \\ Decisions
         & Deciding verdicts (\textbf{10}) & Intuition, Collaboration & `past experience', `multiple perspectives' \\
        \& processes & Verdict-dependent (\textbf{6}) & Subjectivity, Difficult claims  & `conflicting evidence', `explain falsity' \\
         & Communicating & Verdict labels, Adding nuance  & `binary verdicts', `giving more context'\\
         & complexity (\textbf{9}) & & \\
          & \\
         \textbf{RQ2:} & Claim detection (\textbf{6}) & Claim matching, Efficiency &  `social media', `live debate monitoring'
        \\ Using
         & Evidence retrieval (\textbf{8}) & Collecting evidence, Extracting information & `summarising documents', `data cleaning' \\
        tools & Veracity prediction (\textbf{8}) & Detect AI-generated content, Multi-tool usage & `detecting LLM text', `triangulate verdict' \\
         & Communicating  & Editing, Background context & `LLMs for writing', `contextualising story' \\
          & fact-checks (\textbf{4}) & & \\
          & \\
         \textbf{RQ3:} & Explain processes (\textbf{9}) & Human-in-the-loop, Technical knowledge & `verify model', `understanding algorithm' \\
         Explanatory 
         & Explain uncertainty (\textbf{8}) & Noting uncertainty, Unexplained uncertainty & `distrust uncertainty', `uncertainty source' \\
         needs & Replicability (\textbf{9}) & Leave it up to reader, Explaining processes & `readers check themselves', `using links'\\
         & Verifiability (\textbf{9}) & Checking reliability, Signposting evidence & `point to where to look', `citing sources'\\
         & Faithfulness (\textbf{3}) & AI hallucination, Human-like explanation  & `mirror human processes', `inaccurate output'\\
    \hline    
    \end{tabular}
    \caption{Example themes, subthemes and codes developed from analysis of interview transcripts}
    \Description[The table summarizes research questions, themes, sub-themes, and example codes related to automated fact-checking tools.]{The table summarizes research questions, themes, sub-themes, and example codes developed during the interviews  It is organised into columns representing these aspects.
    Column 1 (Research Questions): Three research questions are highlighted:
    RQ1: "Fact-checkers' decisions and processes."
    RQ2: "Uses of automated fact-checking tools."
    RQ3: "Explanatory needs of fact-checkers."
    Column 2 (Themes): Themes under each research question are listed:
    For RQ1: Evidence quality, deciding verdicts, verdict dependent
    For RQ2: Claim detection, evidence retrieval, veracity prediction, communicating fact-checks
    For RQ3: Explain processes, explain uncertainty, replicability, verifiability, faithfulness
    Column 3 \& 4 (Sub-Themes): Each theme is divided into sub-themes, such as:
    For Evidence quality: "Searching for evidence, "Contextual clues," and example codes: "cross-checking," "primary sources,"
    For Deciding verdicts: Intuition, collaboration and example codes: "past experience", "multiple perspectives."
    For verdict-dependent: subjectivity, difficult claims and example codes: conflicting evidence and explain falsity
    For Communicating complexity: verdicts labels, adding nuance and example codes: binary verdicts and giving more context.
    For Claim detection: claim matching, efficiency and example codes social media and live debate monitoring
    For evidence retrieval: collecting evidence, extracting information and example codes: summarising documents and data cleaning
    For Veracity prediction: detect AI-generated content, multi-tool usage and example codes: detecting LLM text, triangulate verdicts
    For communicating fact-checking: editing, background context and example codes: LLMs for writing, contextualising story
    For Explaining processes: Human-in-the-loop, Technical knowledge and example codes: 'verify model’, ‘understanding algorithm’
    For communicating fact-checking: editing, background context and example codes: LLMs for writing, contextualising story
    For Explaining processes: Human-in-the-loop, Technical knowledge and example codes: 'verify model’, ‘understanding algorithm’
    For explain uncertainty: Noting uncertainty, Unexplained uncertainty and example codes: ‘distrust uncertainty’, ‘uncertainty source
    For Replicability: Leave it up to reader, Explaining processes and example codes: 'readers check themselves’, ‘using links’
    For Verifiability: Checking reliability, Signposting evidence and example codes: 'point to where to look’, ‘citing sources’
    For Faithfulness: AI hallucination, Human-like explanation and example codes: ‘mirror human processes’, ‘inaccurate output'
    }
    \label{tab:RQ_codes}
\end{table*}



\end{document}